%% file: main.tex
\documentclass[preprint,journal]{vgtc}        

\ifpdf
  \pdfoutput=1\relax                   
  \usepackage{graphicx}                
  \DeclareGraphicsExtensions{.pdf,.png,.jpg,.jpeg} 
\else
  \usepackage{graphicx}                
  \DeclareGraphicsExtensions{.eps}     
\fi%

\graphicspath{{figures/}{template/}{images/}{./}} 

\usepackage{microtype}                 
\PassOptionsToPackage{warn}{textcomp}  
\usepackage{textcomp}                  
\usepackage{mathptmx}                  
\usepackage{times}                     
\usepackage{cite}                      
\usepackage{tabu}                      
\usepackage{booktabs}                  

\usepackage{dirtytalk}

\usepackage[inline,ignoremode]{trackchanges} 
\addeditor{jd}
\addeditor{mm}

\usepackage[table,xcdraw]{xcolor}
\usepackage[normalem]{ulem}
\useunder{\uline}{\ul}{}
\usepackage{adjustbox}

\usepackage{enumitem}

\usepackage{color}
\definecolor{cb_blue}{rgb}{0.22,0.49,0.72}
\definecolor{cb_green}{rgb}{0.3,0.67,0.29}
\definecolor{cb_orange}{rgb}{0.84, 0.45, 0.06}
\definecolor{cb_red}{rgb}{1.0, 0.0, 0.0}
\definecolor{cb_periwinkle}{rgb}{0.4, 0.2, 0.8}
\definecolor{black}{rgb}{0,0,0}

\newcommand{\new}[1]{{\textcolor{black}{#1}}}

\onlineid{0}

\vgtccategory{Research}
\vgtcpapertype{theory/model}

\title{Criteria for Rigor in Visualization Design Study}

\author{Miriah Meyer and Jason Dykes}
\authorfooter{
\item
 Miriah Meyer is with the University of Utah. E-mail: miriah@cs.utah.edu.
\item
 Jason Dykes is with City, University of London, E-mail: j.dykes@city.ac.uk.
}

\shortauthortitle{Meyer & Dykes: 
Criteria for Rigor in Visualization Design Study
}

\abstract{
We develop a new perspective on research conducted through visualization design study that emphasizes design as a method of inquiry and the broad range of knowledge-contributions achieved through it as multiple, subjective, and socially constructed. From this interpretivist position we explore the nature of visualization design study and develop six criteria for rigor. We propose that rigor is established and judged according to the extent to which visualization design study research and its reporting are
{\fontfamily{qag}\selectfont{\small{INFORMED}}},
{\fontfamily{qag}\selectfont{\small{REFLEXIVE}}},
{\fontfamily{qag}\selectfont{\small{ABUNDANT}}},
{\fontfamily{qag}\selectfont{\small{PLAUSIBLE}}},
{\fontfamily{qag}\selectfont{\small{RESONANT}}}, and 
{\fontfamily{qag}\selectfont{\small{TRANSPARENT}}}.
This perspective and the criteria were constructed through a four-year engagement with the discourse around rigor and the nature of knowledge in social science, information systems, and design. We suggest methods from cognate disciplines that can support visualization researchers in meeting these criteria during the planning, execution, and reporting of design study. Through a series of deliberately provocative questions, we explore implications of this new perspective for design study research in visualization, concluding that as a discipline, visualization is not yet well positioned to embrace, nurture, and fully benefit from a rigorous, interpretivist approach to design study. The perspective and criteria we present are intended to stimulate dialogue and debate around the nature of visualization design study and the broader underpinnings of the discipline.

} 


\CCScatlist{ 
 \CCScat{Human-centered computing}{Visualization design and evaluation methods}%
 \CCScat{Human-centered computing}{Visualization theory, concepts and paradigms}
 \CCScat{Human-centered computing}{Interaction design theory, concepts and paradigms}
}

\ieeedoi{10.1109/TVCG.2019.2934539}

\vgtcinsertpkg

\begin{document}


\firstsection{Introduction}

\maketitle



\input{_introduction.tex}

\input{_relatedwork.tex}

\input{_perspective.tex}


\input{_criteria.tex}

\input{_discussion.tex}

\input{_conclusion.tex}

\acknowledgments{
This work has been informed by lengthy and active discussions with many colleagues in numerous places over a long period.
In particular we
thank Nina McCurdy and Jonas L\"{o}wgren for deeply influential contributions to our thinking. We also thank
Geraldine Fitzpatrick, Ethan Kerzner, Alex Lex, Julienne Meyer, Torsten M{\"o}ller, Michael Sedlmair, Brett Smith, Jason Wiese, Jo Wood and four engaged and widely informed reviewers for provocative critical conversations. 
Katharine Coles, @jamesscottbrown and @antarcticdesign had late effects.

}

\bibliographystyle{abbrv-doi}

\bibliography{template,criteria}

\input{_tableEnd.tex}



\end{document}

%% file: _introduction.tex
\label{sec:introduction}

Design study -- an approach to applied visualization research \cite{sedlmair12} -- is now a standard method for conducting visualization inquiry, guided by validation methods \cite{munzner09,meyer15}, process models \cite{sedlmair12, mckenna2014design, mccurdy2016action}, scenarios \cite{sedlmair16}, and an increasing set of representative examples in the literature \cite{zhang2019idmvis,nobre2019lineage,brehmer2014overview,meyer2010multeesum,meyer2010pathline,kerzner2017graffinity,goodwin2013creative,sultanum2019doccurate,hinrichs2015speculative}.
In the context of the wider visualization discipline that is increasingly assessing its practices, the maturing of design study has exposed a series of provocative, open questions that we hear researchers asking:
\new{
What are the research contributions made through design studies, and do they generalize? 
What is the value of specific solutions?
If a design study is not reproducible, can it be rigorous? 
How do we conduct design study research well, and how do we assess it?
Is design study even research?
}

\new{
Underlying these questions is a strong focus in the community on the production of visualization software systems within a design study \cite{sedlmair16}. Process and decision models used by design study researchers prescribe steps and considerations to design and validate such tools, resulting in a myriad of validated systems. These open questions, however, highlight a problem being faced by researchers seeking to use design study to learn about and express a broader collection of knowledge: process alone does not provide guidance on important considerations for rigor and the construction of diverse forms of knowledge acquired through design \cite{mccurdy2019framework,hinrichs2017defense}. The result over the years has been design studies and their resulting papers that focus on deployed, working software, rather than on taking full advantage of the situated, complex, and nuanced learning that researchers (can) acquire through deep engagement with people, data, and technology. 
}

\new{
In this paper we \say{separate the criteria from the craft} \cite{tracy10} to support the broader set of outcomes that can result from design study research. 
We propose considerations for achieving rigor in, and constructing knowledge through, design study that compliment existing processes.}
We constructed these considerations from a four-year engagement with the ongoing and interrelated discourses around knowledge generation and rigor in social science, information systems, and design. This debate is deep and extensive, and also contradictory, dynamic, and imperfect. It is as complex, messy, and nuanced as the richly situated contexts in which researchers in these fields engage. 

\new{
We explore the theoretical underpinnings of design study and offer a new, interpretivist perspective. This perspective emphasizes design as a method of inquiry into complex, situated, dynamic problems, and the knowledge achieved through it as multiple in form, various in range, and inherently subjective and socially constructed. From this perspective, we propose a preliminary set of six
criteria for rigor in visualization design study that are intended to guide  researchers in constructing, communicating, and assessing rigorous knowledge claims.} 
We explain why each criteria is an important consideration for rigor, and identify methods to augment current practices that might help to achieve them. Our view is that attaining these criteria is challenging and adopting them may require action by the community, so we pose several provocative questions that are intended to explore implications, sharpen views, and stimulate debate about whether and how this new perspective on design study might be achieved.

But before we provoke, we first attempt to persuade. We develop the theoretical backdrop of our position in Section \ref{sec:theoretical-backdrop}, summarizing a range of thinking and debate from the social sciences, information systems, and design on the nature of knowledge and the ways in which it is constructed. We position our perspective on design study against this theoretic backdrop through a series of statements on the nature of design study in Section \ref{sec:research-through-design-study}, and then propose six criteria in Section \ref{sec:criteria} for establishing rigor from this perspective. In Section \ref{sec:questions} we select three debatable questions that our perspective opens up -- there could be many more -- and offer our initial opinions. Finally, we conclude with a call to the community to critique and debate our work, as well as the broader philosophical underpinnings of visualization research.

%% file: _relatedwork.tex
\section{Theoretical Backdrop}
\label{sec:theoretical-backdrop} 

The perspective on design study that we present in this paper is informed by a close reading of literature about rigor in social science, information systems, and design. In this section we provide a brief overview of the main themes and threads of discourse that informed our thinking.

\subsection{Philosophical Positions}

The predominant philosophical position in science, computer science, and visualization is that of \emph{positivism}, which views reality as singular and external, on the basis that it can be objectively known. Positivist research approaches focus on reducing researcher reactivity, and achieving reliability, replicability, and representativeness \cite{burawoy98}. Data are collected and analyzed with the aim of producing an unambiguous result that is representative of the single reality. Validity criteria for establishing the truthfulness of results rely on reproducibility and replication \cite{cook1979design}, the achievement of which underlies many positivist approaches. Active discussions on these issues in visualization research focus on the reproducibility of data, data transformations, interactive exploration, algorithms, and software systems \cite{silva07,haroz2018open}, as well as
the reliability of user studies \cite{sukumar2018towards,crisan2018evaluate}.

Discourse in the social sciences and humanities has a long history of critiquing positivist positions \cite{haraway88,burawoy98,firestone93}, particularly for studies that seek to understand people and their experiences \cite{lincoln85}. These arguments advocate for a \emph{relativist} position that considers reality to be multiple and mind-dependent \cite{Smith2017}, and the researcher as an active instrument of the research. \new{In contrast to the positivist position, the subjective nature of knowledge is a key component and strength of relativist methodologies \cite{finlay02}.} To support a relativist standpoint, \emph{interpretivist} approaches view the knowledge that a researcher acquires as socially constructed -- rather than objectively determined -- and use methods such as dialogical approaches that are spoken, written, and interpreted.
\new{
Subjectivity is embraced and considered shorthand for the construction of knowledge through interpretation \cite{drucker2011humanities}.
}


\new{Visualization research benefits from both positivist and interpretivist approaches as it involves multiple types of phenomena and context.} Many perceptual, cognitive, and computational phenomena can be studied effectively through controlled, empirical studies where objectivity, repeatability, and prediction are valued and efforts are made to remove bias and error. Studying people and their considered, complex, contextualized, social reactions to dynamic settings often benefits instead from relativist approaches that involve subjective interpretation of qualitative data \cite{kerzner2019framework,hogan2016elicitation,sultanum2019doccurate,Walny2018,lee2016people,thudt2017subjectivity,drucker2011humanities}. A key concern, however, is that work that is conducted from one position is judged from another \cite{dourish96}. Positivists might question research that involves subjectivity or bias. Interpretivists, however, are likely to question research rooted in positivism that does not account for the inherently subjective judgments involved in most knowledge construction \cite{drucker2011humanities}. \new{We argue that the visualization community is missing a broadly shared understanding of how research emerging from these very different philosophical positions is undertaken with rigor.}

\subsection{Interpretivist Criteria}			
		
A considerable challenge for interpretivist research approaches is that of establishing rigor criteria that consider the \say{creative complexity of the qualitative methodological landscape} \cite{tracy10}.
In their explicit rejection of the positivist notions of rigor, the seminal work of Lincoln and Guba \cite{lincoln85} established interpretivist criteria for judging the \emph{trustworthiness} of research. The criteria \emph{credibility}, \emph{dependability}, \emph{confirmability}, and \emph{transferability} are offered as alternatives for scientific validity and generalizability to instead consider: \say{How can an inquirer persuade his or her audiences that the findings of an inquiry are worth paying attention to?} \cite{lincoln85}. The underpinnings of, and methods for achieving trustworthiness have been considered, debated, expanded, rejected, and reaffirmed extensively in the literature since \cite{tracy10,guba2005paradigmatic,morse15,smith18,polit10,seale1999quality}.

The difficulty in establishing criteria for qualitative, interpretivist research that is inherently messy, changing, subjective, and context-specific is in stark contrast to the strong consensus for positivist approaches that aim for validity, reliability, generalizability, and objectivity -- this tension is in part responsible for the undervaluing and undermining of qualitative work \cite{tracy10,seale1999quality}. Resolving this tension has led to a proliferation of perspectives on rigor criteria in the literature. For example, Lincoln and Guba subsequently reject much of what was presented in their original work \cite{guba2005paradigmatic}. Morse calls researchers to reclaim the validity and generalizability constructs of positivist criteria in the qualitative realm \cite{morse15} as have others \cite{smith18,polit10}. And Tracy provocatively offers eight universal criteria for quality in qualitative studies \cite{tracy10}, which are much used and routinely critiqued \cite{Smith2017}. 

In this paper we take small steps into the conversation through a proposal of rigor criteria specifically applicable to visualization design study. Our development of the criteria was informed in-part by the work and thinking of these social science scholars, and it follows in their tradition of rejecting wholesale assimilation in favor of more nuanced criteria suited to a specific approach to research.	 	 		

\vspace{2mm}

\subsection{Design Research}

The tensions and synergies within and between design and research are considered in a series of related fields. In the applied field of information systems, scholars grapple with the competing needs of design practice and academic research in the context of developing technology for and within organizations \cite{sein11}. Influenced by \emph{design science} -- a problem-solving paradigm that seeks to analyze, design, implement, and manage information systems \cite{von2004design,hevner2010design} -- Sein et al. propose
guiding principles through their \emph{action design research} \cite{sein11} methodology.
These principles capture a view that the design of information systems should be both guided by a researcher's intent and shaped by the organizational context, which resonates with the goals of design study \cite{sedlmair12}. The principles
are useful for considering the role of people and context in shaping visualization artifacts within design study, and for recording and reporting on these effects \cite{mccurdy2016action}.

The action design research method, and underlying principles, stem more broadly from \textit{action research}, an approach that relies on action as a means for developing knowledge \cite{lewin46,hayes11}. As a democratic approach, action research emphasizes researching \emph{with} people in their everyday, real-world contexts, not \emph{on} them through a process that cycles between planning an intervention, enacting the intervention, observing changes based on the intervention, and reflecting on the changes in order to plan for another cycle. Melrose describes the effects of these cycles on the research process as: \say{the [researchers and participants] make mistakes and learn from them, so the research design and questions are emergent and changeable... [it] is an unrepeatable journey with unpredictable results and undreamed of conclusions}\cite{melrose01}.  
Although action researchers, like other social scientists, question the meaning of rigor for their work, the literature on action research points back to Lincoln and Guba's original notion of trustworthiness \cite{lincoln85}.

An alternative thread of thinking and discourse on the production of knowledge through design comes from the \emph{research through design} (RtD) community. RtD is an approach embraced by design researchers and academics who view design artifacts as experiments in future possibilities and the expression of knowledge a researcher gains about those possibilities \cite{frayling1994research,zimmerman07,lowgren07}. Importantly, RtD emphasizes the production of knowledge by means of design activities \cite{stappers17}. Theoretical work in RtD examines the nature of knowledge generated through design \cite{hook12,lowgren13,cross10,bowers12}, the ways in which design researchers design \cite{duncan04,cross82,cross99,dorst01}, and the relationship of RtD to the goals and values of HCI \cite{zimmerman07,zimmerman10,hauser18,gaver10}.

\new{
Like relativist positions in social science, RtD strongly rejects positivist approaches to research. Instead, researchers argue for the need to embrace a designerly view of knowledge generation that considers the richness and complexity of the design process, context, and outcomes \cite{gaver10,stolterman08,buchanan92,gavers12}. These views place knowledge generation within \say{specific, intentional, and non-existing} design contexts \cite{stolterman08}, and result in particular, situated outcomes that are subject to a designer's unique perspective \cite{gaver10}. Whereas some RtD researchers argue that methodological standards \say{threaten to occlude the potency of unique, embodied artifacts in a cloud of words and diagrams} \cite{gavers12}, others argue for a \say{philosophical and methodological understanding of \textit{what constitutes the rigor and discipline of design practice} in order to better support practice} \cite{stolterman08} (emphasis in original). The synergies between design research and social science have led to recent calls in the RtD community for design researchers to more fully and systematically embrace methodological approaches of the social sciences \cite{frayling15,redstrom19}.
}


\subsection{Visualization Design Study}

\new{
In visualization, \emph{design study} 
\footnote{Visualization design study is not explicitly related to the academic design discipline of design study -- we note that our reference to design study throughout this paper is with respect to the visualization community's definition.
} 
is defined and described as an applied methodology by Sedlmair et al. \cite{sedlmair12}: it is \say{a project in which visualization researchers analyze a specific real-world problem faced by domain experts, design a visualization system that supports solving this problem, validate the design, and reflect about lessons learned in order to refine visualization design guidelines.} This definition requires that visualization solutions are designed for a problem that exists in the world, with domain experts and their data. Through the consideration of literature, observations, interviews, and their own experiences, design study researchers build an understanding of a problem domain and the inherent analysis questions. They operationalize the domain questions into a data representation and set of tasks \cite{fisher2017making}, which guide the design of visualization solutions. Design study researchers purposefully assess their understanding of the domain
and
the efficacy of their operationalizations and visualization solutions through checks with collaborators, data sets, and existing theory and practice. These assessments can be iterative and multiscale throughout the design process; small, rapid assessments are embedded in larger, longer-term ones \cite{mccurdy2016action}. Reflection
, most often 
at the end of a study \cite{2018_cga_reflection}, articulates the learning that occurred to add to the body of visualization knowledge.}

\new{Researchers conducting design studies often employ existing visualization models to guide the methodological structure of the study. Process models such as the nine-stage framework \cite{sedlmair12}, the design-activity framework \cite{mckenna2014design}, or action-design research \cite{sein11,mccurdy2016action} provide guidance for the high-level steps a researcher could take to conduct a design study, with recommendations of specific methods for each step. Complementing these process models, 
the nested model\cite{munzner09} is an often-used design decision model that provides guidance for choosing appropriate approaches for validating a visualization system. This model categorizes visualization design decisions at four levels, and identifies validity threats for a designed visualization system at each.
}

\new{
Several forms of knowledge contribution can be achieved through design studies \cite{sedlmair12}: a characterization of the problem domain, a validated visualization design, and improvements to visualization guidelines. Current guidance emphasizes the importance of reflection in establishing these claims, which, for design study, \say{is where research emerges from engineering}\cite{sedlmair12}. The visualization community, however, has not reached consensus in the about how to reflect, when to reflect, or how to improve and judge the quality of the subjective reflective process \cite{2018_cga_reflection}. The nested model supports \emph{testing} the validity of (some) forms of design study knowledge claims, but it does not provide insight or guidance into reflectively \emph{generating} them. Similarly, design study process models articulate steps to take, but don't articulate \emph{how} to produce meaningful, varied, and valuable knowledge, or \emph{what} the criteria might be to judge the resulting knowledge claims. }

\new{As a result, there is an emphasis in design studies and their resulting papers on the validated visualization design -- working software appreciated by domain experts -- rather than on the situated, complex, subjective, and nuanced learning acquired by visualization researchers through design study.
We argue that methods for developing valid visualization systems are more accepted, expected, and utilized than methods for reflecting on the processes to establish knowledge claims. Thus, in this work we aim to be more explicit about how visualization researchers can assess their decisions in planning, conducting, and reporting on design study; what they can learn through design study; and how others can judge resulting knowledge claims. 
}


%% file: _perspective.tex
\section{Wicked Subjective Diverse Design Study}
\label{sec:research-through-design-study}


In this section we detail a \new{new, interpretivist perspective on visualization design study that extends and deepens the existing definition}. This perspective embraces design as a \new{subjective} method for constructing and communicating new knowledge, assuming multiple and mind-dependent realities. We present this perspective through four statements on the nature of design study informed by our interpretation, synthesis, and application of \new{approaches from both the qualitative social sciences and RtD} described in Section \ref{sec:theoretical-backdrop}. 
For each statement we point to relevant \new{rigor} criteria that we explain and discuss in detail in Section \ref{sec:criteria}.


\subsection*{Design study uses design for inquiry and expression.}
\label{sec:design}
Design study researchers learn through the design process. They design visualization solutions for real-world problems in close collaboration with domain experts in order to approach the problem in possibly new ways, and to learn by doing so. Researchers explore possibilities through broad consideration of design spaces, and express and communicate much of their learning through design instances and artifacts, such as sketches, prototypes, models, and software systems. Therefore, visualization design study aligns with RtD as \say{a research approach that employs methods and processes from design practice as a legitimate method of inquiry} \cite{zimmerman10}. 
In line with RtD, much of what a design study researcher learns about the molding of \emph{materials} -- combinations of hardware, software, data, and possibly physical materials -- into the developing solution, and the relationship of this solution with the problem, is established through \new{the practice of design}. This approach prioritizes finding solutions to a problem through making and prototyping over theoretical reasoning \cite{cross82,stappers07}. What a designer comes to know is frequently expressed implicitly through the design itself: its visual form, its interactive characteristics, and the subtle ways in which materials are shaped to address the problem \cite{cross99}. We consider the same to be true of design study researchers.


Design study researchers are particularly attuned to opportunities for constructing and testing knowledge through design -- ideas, concepts, encodings, interactions, and their combination -- and to engage, observe, and collect appropriate evidence to explore these possibilities \cite{mccurdy2019framework,goodwin2013creative,hinrichs2017defense}. Taking the perspective of a visualization as a \emph{technology probe} \cite{hutchinson2003technology} offers opportunities to learn about the relationship of people and data beyond learning about the visualization itself \cite{mccurdy2019framework}. Ultimately, design study researchers construct knowledge \new{subjectively} through \new{reflective} critical reasoning based upon experience and evidence established through the study, and against a backdrop of existing knowledge. 

Design researchers understand that \say{the whole point of doing research is to ... make knowledge available to others in re-usable form} \cite{cross10}. Visualization
design study researchers predominately make their knowledge available through written reports. They aim to produce explicit and appropriately scoped expressions of  knowledge claims that allow them to be communicated persuasively, effectively, and in ways that resonate with the community. Reports primarily take the form of an academic paper including its prose, figures, and other constituent parts.
Additional forms of effective knowledge expression
include imagery, software, digital artefacts and videos with annotations and narrative.

\vspace{0.1cm}
\new{
Design as a method of inquiry and expression leads us to suggest five criteria for rigor -- that the design process:
is {\fontfamily{qag}\selectfont{\small{INFORMED}}} by existing designs to inspire and understand candidate solutions;
is {\fontfamily{qag}\selectfont{\small{ABUNDANT}}} in observations, designs, and descriptions;
produces {\fontfamily{qag}\selectfont{\small{PLAUSIBLE}}} designs and interpretations of design processes;
generates designs and claims that are {\fontfamily{qag}\selectfont{\small{RESONANT}}};
and expresses knowledge claims explicitly through {\fontfamily{qag}\selectfont{\small{TRANSPARENT}}} description and evidence.
}


\subsection*{Design study tackles wicked problems.}
Design study researchers design artifacts based on their understanding and interpretation of a domain problem. A visualization is thus not only an expression of knowledge, but also a technological representation of a problem expressed through a potential solution. By developing visualization designs in close consideration with domain experts and the context of use -- continually reassessing their form, function, and potential \cite{sein11} -- the design study researcher shifts and shapes the solution to effectively address a problem
that is of interest to domain experts.  

The iterative, dynamic shaping of the problem and its expression through the designed solution illustrates the {\it wicked nature}  \cite{rittel1974wicked} of the problems tackled by design study researchers. \new{Wicked problems are {\it indeterminant}, meaning \say{there are no definitive conditions or limits to the design problem} \cite{buchanan92}.} An important characteristic of wicked problems is that it is \say{only in terms of a conjectured solution that the problem can be contained within manageable bounds.} \cite{cross82}. The problem definition is considered a design space just as the solution is, with progress towards defining one affecting the progress of defining the other \cite{maher96,dorst01,zimmerman14}. \new{Wicked problems have unbounded potential for solutions due to the complexity of design \cite{stolterman08}, the absence of inherent stopping criteria \cite{buchanan92}, and a designer's articulation of the problem and solution as one of many possible interpretations \cite{bowers12}. These solutions cannot be assessed as true or false, but rather as good or bad \cite{buchanan92}.}

\new{Embracing wicked problems as core to design study has several implications.}
First, wicked problems encourage input from both designers and domain experts, shaping designs into solutions that are relevant, meaningful, and interesting. The solutions are inextricably related to the problem, the design approach, and the people, including the design study researcher who is defining both problem and solution in ways that are necessarily interdependent, highly subjective, and fluid. Second, evidence of the changing problem and solution, and their regular shaping and shifting, is an indication of a strong collaboration between design study researchers and domain experts working toward a mutually beneficial solution. Instability of a problem definition, identified through changing focus and expressed through task requirements, is  a measure of success for design study \cite{sedlmair12}. \new{Third, the design of {\it good} solutions requires the consideration of a broad space of possibilities \cite{sedlmair12,stolterman08}. 
}

\vspace{0.1cm}
\new{
The wicked nature of problems that design study tackles requires two criteria for rigor:
An {\fontfamily{qag}\selectfont{\small{ABUNDANT}}} approach to allow multiple voices and perspectives to shift and shape design problems and consider a broad set of solutions;
and that evidence of the dynamic process is reported in {\fontfamily{qag}\selectfont{\small{TRANSPARENT}}} ways.
}



\subsection*{Design study is inherently subjective.}
What design study researchers learn is personal, subjective, and specific. The situated and inherently wicked nature of the visualization design process means that knowledge acquired through design study can only be understood within the context of its construction. This \emph{context} includes not only the views and experiences of domain experts, datasets, organizational and social constraints, but also a design study researcher's own intuition, interests, experiences, and values. The researcher has important effects on the artifacts that she produces, the problems she addresses, the activities and reactions she observes and interprets, and the details and knowledge she chooses to report. Visualizations, and the visualization design process, are not neutral \cite{dork13,correll19}.

Knowledge constructed in this way is inherently interpreted \cite{smith18, Smith2017} and subject to the many assumptions, values, and commitments that researchers bring to their work \cite{braun2013successful}. A relevant position for design study is that the observable world can never be construed devoid of and separate from those that observe it \cite{burawoy98}. Therefore, we argue for a relativist perspective to design study, in which observed realities are accepted as multiple, relative, changing, and mind-dependent \cite{denzin1994handbook, finlay02b, smith18, Smith2017}. This position contrasts with positivist approaches that are prevalent in the visualization research community and assume the researcher to be a distant, objective observer of a singular reality.

Research that takes a relativist standpoint can draw upon established methods to develop meaningful knowledge from deep engagement in, and description of, the context in which the observations and experiences take place.
\new{These methods utilize subjectivity to support a researcher in diversifying the perspectives and views she is studying, to better understand the varied viewpoints of her participants, and to recognize her own learning and construction of knowledge \cite{finlay02}.}
Design study researchers have significant opportunities to adopt appropriate relativist epistemologies from other fields by investing in methods for generating, reporting, sharing, and using constructed knowledge. 

\vspace{0.1cm}
\new{
Four rigor criteria embrace the inherent subjectivity of design study:
understanding and leveraging the role of the researcher through a {\fontfamily{qag}\selectfont{\small{REFLEXIVE}}} design process;
{\fontfamily{qag}\selectfont{\small{TRANSPARENT}}} communication of her effects;
and the development of {\fontfamily{qag}\selectfont{\small{PLAUSIBLE}}} claims from 
observations and analysis that is {\fontfamily{qag}\selectfont{\small{INFORMED}}} by appropriate epistemology.
}


\subsection*{Design study produces diverse knowledge claims.}
The knowledge that design study researchers construct varies greatly in topic, form, and range. In line with the perspective developed here, we define \emph{knowledge} as something a design study researcher comes to know through an inquiry.
Design study researchers' focus on context-informed development and use of technology allows them to learn various things in multiple ways and at multiple scales about:

\vspace{0.05cm}

\begin{itemize}[nolistsep,noitemsep]
\item \emph{visualization idioms}: particular graphical representations of data, how well they support activities in particular contexts, and how broadly they might apply
\item \emph{design guidelines and methods}: effective ways of developing solutions and undertaking visualization design and design study
\item \emph{problem domains}: the relationships between people, data, and technology, situated within a specific domain.
\end{itemize}

\vspace{0.05cm}

\new{
This diversity of \textit{topics} stems from learning acquired through the practice of design as well as through purposeful examination of the existing world. In this way, design study knowledge construction reflects approaches taken in both RtD and the social sciences.}

 The diverse \textit{forms} of knowledge expression in design study vary from the use of words, mathematical notation, and pseudo-code, to diagrams, imagery, and design artifacts. This broad definition of knowledge -- from design and relativist perspectives on knowledge construction \cite{smith18,gaver10} -- implies that a design study researcher produces a knowledge expression \emph{every time} an observation is recorded, a sketch is generated, or code is manipulated. In recording details about a situation and a design solution, such as what is said, what is implied, how an encoding is used, and the choices embedded in a successful design,
 an explicit act of abstraction occurs.  The design study researcher decides which details are meaningful, which are not, and how they will be recorded. Whether jotting down observations, sketching design ideas, or manipulating materials like code, data, or other design media, the
 researcher abstracts details of a situation into a new, interpreted knowledge expression.

An important, and contentious, characteristic of knowledge claims concerns their \emph{range}: the amount of the explainable world to which they apply \cite{siponen18}. The range can be considered an indication of the generality of the knowledge along a continuum from the particular to the general, and everything in between \cite{hedstrom98}. Design study researchers produce knowledge across this range. They produce particular knowledge through the design process that focuses  on the current problem and context. This knowledge is specific and situated, such as a detailed and rich description of a domain expert and the ways she uses a visualization tool. Design study researchers also construct more general knowledge by engaging in the analytical activity of abstracting details of the situated design context into knowledge with a broader range. 

This form of more general knowledge is sometimes termed \textit{theory}. Debates in the social sciences \cite{Smith2017,hedstrom98,boudon91}, information systems \cite{gregor06,siponen18}, and RtD \cite{hook12} literature offer various perspectives on the point at which knowledge becomes theory. The implication that the general is more valuable than the particular is often subtle, but sometimes explicit: \say{We do not regard a collection of facts, or knowledge of an individual fact or event, as theory} \cite{gregor06}. From the relativist perspective, however, even situated, specific knowledge expressions involve theory, as \say{there cannot be theory-free knowledge because a person's understanding of reality is only known through their experiences (i.e., knowledge is socially constructed and thus fallible)} \cite{Smith2017}. Where theory begins on this continuum is contested, but the most important issue, particularly in the design study research context, is the tension between the \textit{explanatory potential} of the general and the \textit{explanatory accuracy} of the specific. Siponen \cite{siponen18} offers nuanced descriptions of valuable theory types across the range, which include \say{grand, wide range, middle range, small range, narrow range, very narrow range, and unique}, and notes that narrowly scoped, particular theories, for example, can have great potential for practical impact.

\new{
Like RtD, design study produces knowledge claims that range from the \textit{ultimate particular} -- the manifestation of a desired reality in a design artifact \cite{stolterman08} -- to the \textit{middle-range} -- a more abstracted concept than a particular instance that does not aspire to the generality of theory \cite{hook12}.
}
A design study paper often presents knowledge claims across this part of the range. For example, in our paper about a design study with energy analysts \cite{goodwin2013creative}, we offered, among others, the following claims, ordered from the most general to the most specific:
\vspace{0.05cm}

\begin{itemize}[nolistsep,noitemsep]
\item The explicit use of \textit{creativity methods} as contributing positively to novel, effective, and well-aligned visualization solutions.
\item The design concept of \emph{data sculpting} for interacting with energy-model outputs through a graphical interface.
\item A software artifact -- called \emph{demand horizons} -- that instantiated the data sculpting metaphor for use in comparing the energy consumption of household appliances and scheduling their use. \end{itemize}

\vspace{0.05cm}
\new{
The construction of diverse forms of knowledge through design study is supported by six criteria for rigor:
{\fontfamily{qag}\selectfont{\small{REFLEXIVE}}} and {\fontfamily{qag}\selectfont{\small{INFORMED}}} practice allows the researcher to recognize learning and develop general concepts;
{\fontfamily{qag}\selectfont{\small{ABUNDANT}}} evidence is used to develop both specific and general knowledge;
and {\fontfamily{qag}\selectfont{\small{TRANSPARENT}}} recording and reporting such that evidence and analysis produce knowledge claims that are both {\fontfamily{qag}\selectfont{\small{PLAUSIBLE}}} and {\fontfamily{qag}\selectfont{\small{RESONANT}}} to the broader community.
}


%% file: _criteria.tex
\section{\new{Six Criteria for Rigor}}
\label{sec:criteria}

Our intention is to develop a set of complimentary criteria that help researchers in making varied, diverse, and appropriate decisions about how to rigorously conduct design study. \new{These criteria, presented in summary in \autoref{tab:criteria} of supplemental materials, are drawn from established criteria and principles in the social sciences and design in support of the interpretivist perspective of design study we present in Section \ref{sec:research-through-design-study}.} We are striving to provide guidance on \textit{what} to achieve in a design study,
and providing suggestions for transferring existing methods from cognate disciplines to achieve this, as opposed to dictating \textit{how} it should be done. 
We leave it to design study researchers to decide, and argue for, how best to achieve these criteria given the specific people, data, and context involved in a study, and in light of the claims that they make and their own research skills and design expertise. 
The six rigor criteria we propose are applicable to many of the current approaches used in conducting design study. \new{We note that it is unlikely, and indeed unnecessary, that a single design study can meet them all. Like designers, design study researchers work within constraints; it is up to the researcher to choose methods and report persuasively, guided by the criteria and informed by the context. It is up to the reader to assess the extent to which a report supports the criteria.}

\new{The development of these criteria spanned a four-year period of deep engagement with literature about rigor in the social sciences, and to a lesser extent the literature from design. Our investigations began with our discovery of the action design research methodology \cite{sein11}, which resonated with our experiences of conducting design study \cite{mccurdy2016action}. This methodology is grounded in seven principles, but ultimately relies on Lincoln and Guba's notion of trustworthiness \cite{lincoln85} for establishing rigor. As we struggled to pragmatically understand the four trustworthiness criteria -- credibility, dependability, confirmability, and transferability -- and relate them to design study we engaged with the extensive debates in the social science literature surrounding rigor, trustworthiness, and the tensions between realist and interpretivist perspectives on knowledge. Interpretivist perspectives pointed us to Tracy's eight \textit{big-tent} criteria \cite{tracy10}, which provide an extended, and updated view on universal criteria for quality in interpretivist research. The extent of the debates on rigor also encouraged us to develop bespoke criteria for design study informed by these perspectives but specific and targeted to the research approaches, views, and culture of the visualization community. Through an iterative, dialogic process of defining and redefining existing criteria, applying them to our work and that of colleagues, our redefinitions slowly took on new views of rigor, settling into the set of six inter-related criteria presented here. In Table \ref{tab:criteria} of supplemental materials, we list the points of reference from the action design research principles, trustworthiness criteria, and big-tent criteria that informed each proposed criterion for rigor in design study research.}



\vspace{0.3cm}

\subsection{INFORMED}
\hspace{0.25cm}{\fontfamily{qag}\selectfont\small{\textbf{{
Existing knowledge informs design and facilitates new interpretations.
}}}}\vspace{2mm}

It is important to approach design study with a \emph{prepared mind} \cite{furniss11}, that is, with a broad awareness of visualization idioms, design guidelines and methods, and assessment techniques \cite{sedlmair12}; the disciplinary underpinnings of visualization -- core topics and area boundaries, ontological and epistemological positions, socio-cultural views \cite{rode11,attia17}; and relevant design materials like datasets, code, software, hardware, and physically manipulable materials.
Existing knowledge serves as a backdrop against which to grapple with and make sense of the design challenge and the research inquiry, from providing a broad consideration space for designing new solutions, to selecting appropriate methods and for interpreting the nuances of the situation under study.  Researchers make sense of the things they observe and experience through relations to existing knowledge, helping to identify and connect the things they learn as they construct new knowledge \cite{rode11,attia17,burawoy98}. Research informed by existing knowledge supports interesting, meaningful, and new insights and interpretations. 

Besides preparation based on existing literature about visualization, such as textbooks \cite{munzner2014vad} and academic papers, design practice emphasizes the value of an extensive repertoire of examples to serve as a library of ideas and knowledge from which to pull \cite{hook12,cross99,bowers12}. A designer uses this knowledge-base when considering options for creating visualizations: the broader the base the more likely the designer is to consider good ones \cite{sedlmair12,schon1987educating,hook12,dow10}. Designers draw on existing knowledge not only to directly transfer relevant techniques and methods to the problem at hand, but also to facilitate ideation on possibly new approaches. 

The abstracted nature of some knowledge, however, contrasts with the specificity of designing a visualization artifact in a particular context. A  guideline like \say{spatial encoding is most effective} \cite{cleveland1984graphical,heer2010crowdsourcing} cannot inform all the design details of a complex, visualization technique like a curvemap \cite{meyer2010pathline}. Visualization idioms such as overview+detail or the treemap cannot dictate how to specifically balance the competing needs of designing for complex datasets that provide a partial picture of more complex phenomena, diverse participants, and real-world analysis tasks. Design methods cannot (and should not \cite{Bigelow2014,stolterman08,gaver10}) inform every step that a designer takes when ideating, making, and assessing. 
\new{Instead, through a \textit{preparation for action}, a designer \say{acts on a situation with a regard for all of its richness and complexity, and in a way that is appropriate for the specifics of that situation} \cite{stolterman08}. Through careful consideration of the context of a situation against a backdrop of existing knowledge a designer shapes the design process and artifacts into new knowledge expressions that are \say{informed by current theory, creating an ongoing dialog between what is and what might be.} \cite{zimmerman14}.}

\vspace{0.3cm}

\subsection{REFLEXIVE}
\hspace{0.25cm}{\fontfamily{qag}\selectfont\small{\textbf{{
We effect the research, and it effects us.
}}}}\vspace{2mm}

Embracing the subjective nature of much of the knowledge established through design study is essential given its situated, collaborative, and wicked characteristics. Doing so requires a reflexive perspective that \say{embraces not detachment but engagement as the road to knowledge} \cite{burawoy98}. \emph{Reflexivity} \cite{tracy10,finlay02} is explicit and thoughtful self-awareness of a researcher`s own role in a study, grounded by the perspective that observed realities are multiple and constructed \cite{denzin1994handbook, finlay02b, Smith2017, smith18}. While positivism values objectivity and detachment, interpretivism values a focus on how we actively construct our knowledge \cite{finlay02b}. Reflexivity addresses this by identifying how a researcher's own biases and motivations shape the research process, influence participants and the observations she makes, and how the research process changes her own thinking and actions. Reflexivity accounts for, and leverages, the inherent subjectivity of design study researchers.

Over the years, reflexivity in qualitative research has broadened and evolved \cite{finlay02}, with a number of uses of reflexivity that are particularly relevant for design study: identifying a researcher`s assumptions, reactions, emotions, and blind-spots; developing empathy to enable a researcher to see perspectives other than her own \cite{dadds08}; and recognizing moments of learning that can form the basis of more general concepts. Many existing methods for encouraging, supporting, and reporting reflexive practice developed in the social sciences are candidates for useful application to design study research.

Before beginning a study, reflexive researchers can assess both their readiness as well as their biases through reflection on questions such as: Why am I doing this research? Am I prepared? What are my expectations and assumptions? What is my moral and ethical stance towards the situation under study? As the study proceeds, a reflexive researcher continually examines her own impact on and within the study, and is sensitive to other participants' responses to them, as well as to the voices and other sources of information that may be missing. Perspectives from critical theory \cite{dork13}, feminism \cite{dignazio16}, and broad consideration of the ethics of the research \cite{correll19} provide many other important reflexive considerations. Reflexive examinations can help ensure shaping of both the design solution and the problem by pointing to opportunities for input from other perspectives \cite{sein11,mccurdy2016action}. Reflexivity is encouraged throughout the research process, and can be achieved via observation, reflection, note-taking, discussions with colleagues and participants, and open, authentic accounts in reporting. Autoethnography \cite{duncan04,dillingham2012design} is a specific approach that applies reflexive investigation to self-observation. 

\emph{Reflexive notes} -- in the form of a dedicated diary \cite{attia17} or as additions to field notes -- are important as both a tool for supporting reflexivity as well as for record-keeping of insights, impacts, and decisions regarding design and research. These notes are a way to capture a researcher`s impressions of a situation; her own emotions, responses, hunches, and surprises; and her self-dialogue about ethical considerations and concerns, such as power dynamics and hidden voices \cite{correll19}. Often written as a first person account, they remind the reader of the presence of the researcher within the study. These first-person accounts also emphasize a self-as-instrument perspective, which can allow researchers to use their own experiences and tacit knowledge as a valid source of data to support interpretation of a situation \cite{tracy10}. 

Discussions with colleagues are another valuable tool for reflexivity. Engaging a \emph{critical friend} puts the researcher in \say{a critical dialogue, with researchers giving voice to their interpretations in relation to other people who listen and offer critical feedback… to encourage reflexivity by challenging each others` construction of knowledge... providing a theoretical sounding board to encourage reflection upon, and exploration of, multiple and alternative explanations and interpretations as these emerge
in relation to the data and writing} \cite{Smith2017}. Through the solicitation of feedback from a critical friend, a researcher confronts the understandability of her descriptions, the soundness of her reasoning, and the depth of her interpretation. Additionally, alternative views offered by a critical friend are a reminder to the researcher that for every constructed view, there are other, alternative interpretations \cite{wolcott1994transforming}. \emph{Design critiques} -- where a designer engages with other designers through a critical, public dialog about a design artifact -- serve a similar, critical function for examining the merits of design artifacts \cite{brath16}.
						
Additionally, expressing and explaining constructed knowledge to colleagues through formal structured presentations -- whether early in the process when presenting work-in-progress, or later in the form of a pre-paper talk  -- provides opportunities for critical-friend feedback as a researcher is actively constructing knowledge expressions. The review (and sometimes rebuttal) processes for academic papers are another form of critique that can encourage reflexive thinking on the part of the researcher.
In short, seeking continual, critical feedback from a variety of sources can activate reflexivity throughout a design study.

\vspace{0.35cm}

\subsection{ABUNDANT}
\hspace{0.25cm}{\fontfamily{qag}\selectfont\small{\textbf{{
More is better.
}}}}\vspace{2mm}

Design study is valued for the rich, complex, and varied nature of the contexts in which researchers generate knowledge. This calls for the study itself \say{to be at least as complex, flexible, and multifaceted as the phenomena being studied ... it takes a complicated sensing device to register a complicated set of events.} \cite{tracy10}. A design study with abundance has rich details; many voices, datasets, contexts, and designs; and significant time in the field. Abundant data that is rich in details and varied in perspectives, supports multiple meaningful interpretations that are nuanced and situated. And a design process that is shaped by many perspectives \cite{sein11,mccurdy2016action} and emerges from many tested alternatives \cite{dow10} is likely to lead to better designs. Questions about abundance a researcher can ask include: Did I spend enough time to gather meaningful and diverse data? Are there enough data and detail to support meaningful claims? Did I consider enough design alternatives to justify the visualizations? Are diverse voices used to shape the design and interpretations? Abundance provides opportunities to uncover, relate, understand, and justify meaningful insights.

How much is enough? Answering this question is complex and reliant on the context and constraints of any individual study; however, a prevalent theme in the qualitative research literature is when a researcher reaches \textit{saturation}. Saturation occurs when adding more data, perspectives, designs, or contexts leads to no new insights. Even though pragmatic advice on how to reach saturation is scarce, \say{explaining what saturation means within the context of a study is essential} \cite{bowen08}. 

One important research tool that is used widely in interpretive research and that can contribute to abundance is \textit{thick description}, which is \say{the researcher's task of both describing and interpreting observed social action (or behavior) within its particular context} \cite{ponterotto06}. Thick descriptions themselves are an \say{in-depth illustration that explicates culturally situated meanings and abundant concrete detail} \cite{tracy10}. They consist of rich, nuanced, and detailed accounts of observed actions and the intentionality of those actions, as well as the thoughts, feelings, and responses of participants to those actions.  In contrast to thin description, which aims to report observations independently of intentions or context, thick description is purposefully interpretive rather than explanatory \cite{denzin09}. Thick descriptions are important for design study research as they provide rich evidence of the specifics of a situation -- whether it be of visualization usage, a reaction to a design possibility, a social interaction among group members, or the logging of an insight \cite{tolmie16}. These situations can involve a researcher observing others, but they can also include a reflexive researcher's own thoughts on, feelings about, and behaviors within the study. This rich, personal evidence supports interpretation and abundant reporting of experiences that allow the researcher and others to construct more general knowledge.
			
Abundant, diverse, and detailed data collection benefits from long-term, sustained collaboration \cite{lincoln85,shenton04}, as well as from participatory \cite{Muller1993} and co-design \cite{sanders05} methods. These approaches build trust, develop agency, and invite interest in the design process, supporting deep engagement between the designer and the domain experts \cite{lloyd2011human}. We have known this engagement to help reveal meaningful contextual insights that can shape the design, at times in profound ways \cite{goodwin2013creative}, as well as to provide nuanced insights into the situation under study \cite{mccurdy2019framework}. \textit{Creative visualization-opportunity workshops} \cite{kerzner2019framework} are a specific participatory method that can support \emph{multivocal} abundance, which is the incorporation of multiple and varied voices in the design process to include viewpoints that diverge from those of the majority or with the researcher herself. These workshops also encourage an abundance of ideas for problems and solutions through rapid divergence exercises.	
 		
Rapid, and parallel \cite{goodwin2013creative}, prototyping supports shaping of the design through an abundance of perspectives and constraints. Quickly trying multiple ideas encourages generative thinking \cite{dow10}, while creating opportunities for participants to provide feedback that shapes a design. Zimmerman points to a need to record these \say{design moves, the rationale for these moves, and how different hunches did and did not work out} \cite{zimmerman10}. Thick description provides a compelling medium.

More is better in many ways, but more may also be more difficult to record, process, relate, and communicate. It may take more resources and more time. It may disrupt the creative design process \cite{dillingham2012design}. While thick description of design decisions and their rationale may provide important information, collecting, creating, and recording more ideas, more data, more explanations, and more designs is a threat to the design process and the effective synthesis of knowledge. Efficient methods for thorough recording in ways that are not disruptive but manage a rich, abundant, and diverse evidence base are essential. Literate visualization \cite{wood2019design} is one approach that aims to make design exposition an integral and efficient part of the visualization design process, with flexible templates for supporting reflexive practice and structured documents that aim to address the data management issue. 


\vspace{0.35cm}

\subsection{PLAUSIBLE}
\hspace{0.25cm}{\fontfamily{qag}\selectfont\small{\textbf{{
Knowledge claims are evidenced, appropriate, and persuasive.
}}}}\vspace{2mm}

\new{
Plausible knowledge claims made in the context of subjective, interpretivist inquiry are supported by sufficient evidence that is compared, related, combined, and linked in appropriate ways. They are expressed explicitly and persuasively, and they are built coherently from evidence through sound justifications.}
That is, observations are representative of the phenomena and processes under study, and interpretations -- including those that are expressed through a design concept or artifact -- are detailed, thorough, coherent, and congruent with what is experienced, observed, and reported.
Constructing and reporting plausible knowledge claims requires researchers to provide abundant and complementary evidence that is structured and presented in coherent ways.
Plausible knowledge claims give others the confidence to use them.

The plausibility of \emph{particular} knowledge claims -- \emph{a participant used a visualization in this way} -- is heavily reliant upon a researcher's use of appropriate design processes and methods of data collection and analysis. 
\new{
Thick description, reflexive notes, and careful curation of design artifacts support the recording and reporting of plausible, particular knowledge claims.
Interpretivist approaches call for an explication of a researcher's subjective perspective, such as her point of view, experiences, values, and biases that influence the way she interprets the world and constructs an understanding of it \cite{drucker2011humanities}, emphasizing the interrelationships between plausibility and reflexive knowledge construction.}

\vspace{0.2cm}

The plausibility of more \emph{general} claims -- \emph{this is a meaningful visualization design concept} -- instead relies on reflexive, analytical knowledge construction. Researchers generalize from specific, situated details to broader, more abstracted concepts through a process of \emph{analytic generalization} \cite{firestone93}. Polit \& Beck describe the process as being one where the researcher \say{distinguishes between information that is relevant to all (or many) study participants, in contrast to aspects of the experience that are unique to particular participants... [it] is a matter of identifying evidence that supports that conceptualization} \cite{polit10}. 
A predominant method for interpretation and knowledge construction in design study is \emph{reflection}: \say{a process by which experience is brought into consideration . . . to achieve meaning and the capacity to look at things as potentially other than they appear} \cite{brockbank2003action}. The nine-stage framework for conducting design studies emphasizes reflection as a crucial activity \cite{sedlmair12}, but pragmatic guidance in the visualization literature for how and when to reflect is sparse, with many variations in reflective practices and expectations for documentation among design study researchers \cite{2018_cga_reflection}. 

\vspace{0.2cm}

An interpretivist approach for encouraging and recording reflection is that of \emph{memo writing}. Memos are informal analytic notes that \say{catch your thoughts, capture the comparisons and connections you make, and crystallize questions and directions for you to pursue} \cite{charmaz2006constructing}. While a researcher engages with her materials and observations, memo writing can spur new ideas and insights through a (reflexive) conversation-with-self. Careful tracking of memos with underlying data can form a rich account of a researcher's reflective analytic process and reasoning, offering evidence in support of the plausibility of a broad range of knowledge claims. Similarly, explicit recording of design decisions and their justifications through literate visualization supports reflection, interpretation, and analysis while also documenting the process \cite{wood2019design}. 

Using multiple forms of analysis with a diverse set of rich and reflexive observations strengthens the plausibility of general knowledge claims. A design study researcher might combine contextual interviews \cite{holtzblatt1993contextual} with a workshop \cite{kerzner2019framework} to establish an initial understanding of a problem domain, using reflective transcriptions \cite{mccurdy2019framework} to interpret the interviews and open-coding to analyze the outputs of the workshop. She might then ideate on possible visualization solutions through rapid prototyping of storyboards \cite{greenberg2011sketching}, refine her ideas based on feedback gathered through speed-dating \cite{hanington2012universal}, and reflexively develop a visualization software artifact using literate visualization \cite{wood2019design}. Through critical reflection on the artifact against the backdrop of existing visualization idioms, she develops a persuasive justification for, and rich description of a new visualization technique, appropriately scoped to address some, or all of the problems identified. Her report includes both a thick description of how people used the visualization tool and an annotated demo of the system.

\vspace{0.2cm}

\new{
A diverse approach to design study has similarities with \emph{crystallization} \cite{ellingson2009engaging}. This approach from the social sciences involves contrasting and synthesizing multiple observations, contexts, types of data, methods -- which may even rely upon different theoretical frameworks -- to understand and present a complex situation under study. The goal of crystallization is \say{not to provide researchers with a more valid singular truth, but to open up a more complex, in-depth, but still thoroughly partial, understanding of the issue.} \cite{tracy2012qualitative} }

\vspace{0.2cm}

There is no one way to develop plausible knowledge claims. Design study researchers can, and should be flexible and creative with the methods they employ, while being mindful that the methods and representation practices are coherent with the study's goals and constraints. 


\subsection{RESONANT}
\hspace{0.25cm}{\fontfamily{qag}\selectfont\small{\textbf{{
The research inspires understanding and invites action.
}}}}\vspace{2mm}

Research emerges from design study when it \say{adds to the body of knowledge and allows other researchers to benefit from the work} \cite{sedlmair12}.  Research that resonates inspires and affects researchers, designers, practitioners, and others who might use the knowledge. It moves, educates, challenges, and changes them, eliciting deeper understanding, empathy, and knowing. Two specific mechanisms with which design study research can meaningfully impact a visualization audience is through \emph{transferability} and \emph{evocative reports}. Not every design study must resonate in the same way, but like any interpretative research, all high-quality design studies must have impact \cite{tracy10}. Resonant research impacts others in the world.

\emph{Transferability} can occur when a reader believes that the situation under study overlaps in meaningful ways with her own; supporting, motivating, and inspiring a transfer of knowledge from one context to another \cite{lincoln85,shenton04,firestone93}. For example, in a design study with neuroscientists, we found similarities in our challenge of gaining consensus with those described in a paper about a design study with energy analysts \cite{goodwin2013creative}. In that paper the authors detailed their use of a workshop to overcome this challenge -- we transferred this method to our context, and found it to be effective for working with our collaborators as well \cite{kerzner2017graffinity}. 

To support transferability, researchers need to abundantly 
and reflexively 
describe experiences, observations, and abstracted concepts with detailed descriptions and interpretations so that readers can determine similarities to and differences from their own contexts \cite{smith18}. Using prose, such as thick description, is one effective way to do so. Both researchers and readers \say{share a responsibility when it comes to assessing the value of a particular set of qualitative research findings beyond the context and particulars of the original study} \cite{chenail2010getting}. When evaluating the potential for transferability, a reader can ask: Do the relevant characteristics of the study's context remind me of others? Researchers can enhance transferability through rich descriptions of the relevant characteristics of the context that seem important contributors to the knowledge they are claiming. These descriptions facilitate readers in making judgments about which contexts are similar enough to transfer \cite{polit10}. 
\new{The use of familiar data sets and data abstractions when demonstrating a new visualization design is likely to facilitate transfer.}

\emph{Evocative reports} inspire new understanding, empathy, and action. Inspiration stems from research reports and design artifacts that encourage the audience to feel, think, interpret, react, or change. Tracy suggests that \say{like a good song or good piece of pie, an [evocative] qualitative report is not boring. It surprises, delights, and tickles something within us.} \cite{tracy10} We have experienced this kind of delight with various good pieces of a visualization pie: the wobbly topography demos provided by Willett et al. in support of their \emph{lightweight relief shearing} technique \cite{willett2015lightweight}; the fluid interactions and elegant encodings used in the \emph{UpSet} tool \cite{lex2014upset}; the rich, thick descriptions of participant reactions to a personal data system in their homes \cite{tolmie16};
and Georgia Lupi's expressive designs of Kaki King's music shown during her capstone talk at IEEE VIS 2017 \cite{lupi2017data}. 
Sedlmair further suggests that a good problem with which readers can associate makes the value of the design study clear \cite{sedlmair16}.
Although the predominant focus in qualitative research is on evocative written reports that require creative, complex, and beautiful prose, design study offers additional evocative reporting media like design artifacts that invite interaction and engagement, narrated videos, data stories \cite{riche2018data}, and other forms of imagery, annotation, and narrative.


Rich details make an evocative report resonate, just as they allow a knowledge expression to transfer. These details could be thoroughly interpreted descriptions of the research context and observations, the full implementation and realization of a design concept into an artifact, or a portfolio of annotated designs.
\new{
In RtD, \textit{annotated portfolios} support transfer in evocative ways by communicating a designer's subjective, abstract view of what is interesting across a portfolio of designs.
Annotations 
\say{modestly and speculatively reach out beyond the particular} \cite{bowers12} by highlighting similar design characteristics, considerations, and contributions. 
Explicit links between the annotations and design particulars encourage others to interpret and transfer alternative ideas.
}
In contrast to reports of controlled studies, which strive for precision and neutrality, it is the opportunity for resonance through rich, evocative, reflexive, and situated details that makes design study and other interpretive research approaches so compelling.

\subsection{TRANSPARENT}
\hspace{0.25cm}{\fontfamily{qag}\selectfont\small{\textbf{{
The reporting invites scrutiny.
}}}}\vspace{2mm}

A design study is not reproducible, nor should it be. The relativist perspective of design study considers knowledge to be constructed by the researcher, and to present one of many, plausible mind-dependent realities. Neither observations nor the interpretations derived from them reproduce. Even the same researcher may not develop the same interpretations were she to go through the same process. This diversity in potential research outcomes is also a characteristic of RtD where \say{there is no expectation that others following the same process would produce the same or even a similar final artifact} \cite{zimmerman14}. Knowledge claims from design study are not subject to the positivist notion of reliability: \say{applying reliability criteria to qualitative research is incompatible with the belief that theory-free knowledge is unachievable and that realities are subjective, multiple, changing, and mind-dependent} \cite{Smith2017}. Instead, judgements about the quality of the research need to be made in the context of what was done, how it was done, and why it was done. Transparent descriptions of activities, processes, evidence, and claims allow judgments about the other criteria to be made.

\vspace{0.1cm}

Where knowledge is \textit{particular} and constructed through interpretation, richly abundant and reflexive descriptions of observations, experiences, and the knowledge construction process enable readers to make judgments about their plausibility. These need to be documented in a manner that is thorough and findable, and  invites scrutiny.
For design artifacts -- an important form of particular knowledge -- transparency poses an additional challenge: \say{much of the value of prototypes as carriers of knowledge can be implicit or hidden. They embody solutions, but the problems they solve may not be recognized.} \cite{stappers07} While knowledge about the specific visual and interaction techniques embedded within a visualization is visually accessible, the domain problem the visualization is meant to address is not. The RtD community has called for the need for explicit descriptions of this hidden, implicit knowledge to support its transfer \cite{cross10,zimmerman14,stappers17,stolterman08}. For visualization, a \emph{data and task abstraction} \cite{munzner09} is one established way of communicating implicit knowledge ingrained in a visualization artifact: it is an operationalization of the problem that the visualization is meant to solve. Reporting a data and task abstraction is thus one way of increasing the transparency of particular knowledge that is embedded in a visualization artifact. 

\vspace{0.1cm}

For more \textit{general} knowledge that is constructed from these particulars, the quality of the analysis through which generalizations are achieved is central to plausibility claims. Transparent reporting of the analysis and supporting evidence that provides a sense of verisimilitude and vicariousness enables readers to better determine \say{if the findings ring true} \cite{shenton04}. Furthermore, providing transparent access to underlying evidence allows others to perform different analyses, potentially in combination with evidence from another study, to construct different insights, interpretations, and knowledge.
Transparent reporting should be self-critical and include errors, failures, analytical dead ends -- the joys and mistakes \cite{tracy10} of the research process -- with sincerity and frankness. 

\vspace{0.1cm}

Reports on design study are typically achieved through an academic paper \cite{sedlmair12}, where researchers document their process, their designs, and their evidence in support of knowledge contributions. These fixed-length reports, however, are not \say{friends} of thick description \cite{polit10} and richly reflexive details \cite{finlay02b}. Design study papers can be, and increasingly are, supplemented with additional materials that allow for richer details for supporting knowledge claims, including images, software, videos, observations, analysis, field notes, reflexive notes, and audit trails. With increasing amounts of supplemental materials, however, a tension around transparency builds as making the materials navigable, searchable, and interpretable becomes increasingly difficult. But, academic papers are just one way to report knowledge -- might there be others for design study?

\vspace{0.5cm}

We highlight several transparency considerations for each of the other criteria:
\vspace{0.2cm}

\begin{itemize}[nolistsep,itemsep=0.2cm]
  \item {\fontfamily{qag}\selectfont{\small{INFORMED:}}} justify design decisions with respect to existing knowledge; explicate the theoretical, ontological, and epistemological stance of the research
  \item {\fontfamily{qag}\selectfont{\small{REFLEXIVE:}}} disclose reflexive notes and processes
  \item{\fontfamily{qag}\selectfont{\small{ABUNDANT:}}} make a rich body of data and evidence available, findable, and interpretable
  \item {\fontfamily{qag}\selectfont{\small{PLAUSIBLE:}}} provide clear, open, and honest descriptions of analysis processes; release memos, design expositions, and other reflective documents; report on dead-ends and failures
  \item {\fontfamily{qag}\selectfont{\small{RESONANT:}}} communicate implicit, hidden knowledge ingrained in artifacts to support transfer; use problems, datasets, designs and narrative that speak to readers.
\end{itemize}
\vspace{0.2cm}
These considerations emphasize the importance of transparency as a criterion to embrace throughout the  design study, not just at the end.

\vspace{0.25cm}

%% file: _discussion.tex
\section{Questions For Debate}
\label{sec:questions}


In this paper we offer a perspective and set of criteria for visualization design study. This perspective emerged from our own experiences, wide reading, and ongoing discussions with each other and the broader community.
Few of the ideas we present are new in themselves. But in combination, the selection of criteria we present, the identified means of supporting them, and the perspective they construct, point to new approaches and opportunities for research inquiry conducted with design study. 
We acknowledge that this perspective has implications, and many open questions remain.
Indeed, it is our intention to challenge perspectives, stimulate debate, and expose some of the implicit assumptions and practices associated with  design study.
In this section we draw attention to some of these implications with open-ended questions that emerge from the perspective and criteria we propose.

\subsection*{Is our field prepared for rigorous design study?}

\textbf{Probably not}. Visualization courses and textbooks tend not to discuss various ontologies and epistemologies or train visualization researchers in a broad range of methods for design and relativist inquiry \cite{rees2019survey}. Furthermore, design study already demands a very broad base of skills and knowledge -- the abilities to code, wrangle digital data, create, design, and participate effectively in collaborations. Adding the skills and knowledge required to capture, record, and make sense of rich qualitative data may be a stretch for many. Making informed judgments about a design study as readers and reviewers will also require knowledge about these methods and their underpinnings, as well as an openness to interpretivist perspectives. Additionally, our current \textit{paper + supplementary} structure and condensed review times act against the effective generation, exposition, and consideration of the kind of design study we advocate.    

\new{
Embracing a rigorous, interpretivist approach to design study, however, is worth it. Design study can produce a diverse range of knowledge about the complex, messy, nuanced, and evolving relationships of people with data and technology. These insights are important for guiding visualization innovations that are relevant to the changing ways in which people create and consume visualizations in the world. Furthermore, design study provides a means to deeply explore and understand how people relate to data, how data can (and cannot) positively influence decisions, and how data are changing society. We need rigorous research that considers the human-side of data science; visualization design study is one way to do so.}

\new{
We do, however, recognize that advocating for more work, more skills, more time, and more rigor threatens to undermine the now wide-spread acceptance of design study as a valuable means of visualization research inquiry. Our intention is not to make rigorous design study unattainable, but to move the community towards a discussion about what the gold standards might be.  Standards for rigor provide an opportunity to develop new, more efficient approaches to design study by allowing us to better understand the trade-offs of various methods. 
Furthermore, all studies have flaws and limitations; we need criteria for rigor in order to understand them in the design study context to both improve the research inquiry and guide the review process.
}


Visualization venues and forums increasingly  support the discussion of research perspectives, such as the one that we advocate for design study.
Related debate at and around the tutorials, panels, keynotes, and workshops held at IEEE VIS in recent years have partly inspired this work and the questions that we pose at the outset.
As the visualization community continues to expand and diversify, we encourage colleagues to consider these perspectives, continue these discussions, clarify and develop their theoretical and philosophical underpinnings, read, reach out, and engage in the debate.
Doing so will help us better understand, achieve, support, and assess rigor in design study.

\subsection*{Are these criteria enough?}

\textbf{No}. We offer criteria for planning and making judgments about the rigor in design study. But is rigorous work necessarily of high quality? To this question we resoundingly answer: No! 

Tracy's big-tent criteria for high-quality qualitative research~\cite{tracy10} heavily influenced our proposal. While some of these big-tent criteria are captured by those in our list, a number of them are not. Tracy argues that \textit{high-quality} qualitative work must additionally consider criteria that capture the worthiness, relevance, timeliness, significance, morality, and practicality of the research topic, as well as the ethical stance of the research itself. 

Several ethical considerations that relate to design study but are not yet deeply considered by the visualization community are the procedural ethics associated with working with and focusing on other people; ethical issues around funding sources and project focus; and the ethics of exit. This latter consideration raises many interesting questions, but is rarely addressed directly in design study research. Do we leave the field in a manner that has improved the knowledge, capabilities, and capacity of those with whom we work? Are these the most important factors for participants? Is design study research sustainable and beneficial post-study from our collaborators' perspectives? Emerging efforts to develop perspectives on the ethics of data visualization \cite{dignazio16,correll19, dork13} offer important, initial considerations for what we consider to be some of the missing criteria for high-quality design study.

\subsection*{Does our field have adequate disciplinary underpinnings?}

\textbf{No}.
During the four years we spent constructing the knowledge and ideas for this paper, we were continually surprised and excited about the depth of theorizing and degree of debating that occurs in fields related to, but outside, visualization. Every set of literature we reviewed about methodologies, rigor, and knowledge pointed us to new issues and complexities that we had not yet considered. As we addressed the challenge of saturation we began to wonder: Where is \textit{our} disciplinary discourse on the philosophical and theoretical underpinnings of visualization?

Within the visualization community, these discussions happen predominantly in informal and ephemeral venues, such as workshops, panels, and magazine articles. This makes understanding and collating the views of the community on such issues as the ontological and epistemological positions of our work, or the nature of visualization theory, a real challenge. Our main publishing venues need (imperfect, provocative, challenging) papers that openly discuss and critique these sorts of ideas \cite{bongshin2019broadening}. These papers are likely difficult to publish under current reviewing standards as they present opinions and positions that are easily debatable and inherently fallible. We offer this work as one such example.

Gregor suggests four questions that arise from considering the knowledge and theories that characterize a discipline \cite{gregor06}: 1) What are the core problems and interests, and what are the boundaries? 2) What is theory, how is it composed and expressed, and what does it contribute? 3) How is knowledge acquired, tested, and assessed? 4) What are the socio-political considerations, including ethics, power, inclusion, and degree of consensus on these questions?
The visualization community needs to discuss these questions -- publicly, critically, and frequently.

%% file: _conclusion.tex
\section{Conclusion}

As a discipline, visualization has foundations in realism; positivist approaches have enabled us to build an extensive knowledge base about machines, algorithms, and human responses to graphical stimuli.
These approaches have contributed to the development of process models \cite{sedlmair12,mckenna2014design,mccurdy2016action} for studying people and their long-term problem-inspired interactions with data, and decision models \cite{munzner09,meyer15} that guide us to desired outcomes, such as valid visualization software. The full range of knowledge that we might construct through these deep, complex, sustained, situated collaborations, however, is not currently well supported by the predominant approaches, standpoints, and expectations within our discipline. We want to move away from the situation where the kind of knowledge that can be constructed through applied visualization research is questioned because descriptions and context are specific; because applying the same process does not produce the same result; because knowledge is preliminary rather than definitive; because the research involves the subjective voice; because the researcher shifts and shapes the context under study.
In short the kind of richly situated and nuanced knowledge that we can establish through design study is incompatible with positivist positions.

The perspective developed in this paper addresses this concern by embracing relativism.
It is based upon our synthesis of a wide-body of work in related disciplines: we read literature, attended conferences, and engaged with members of these communities. The thinking in these fields on the nature of knowledge and the ways it is constructed in a whole range of contexts 
through inquiry and design is overwhelming in its extent and impressive in its depth. It is also consistently contradictory, surprisingly dynamic, and gloriously imperfect. The thinking will never be complete.

We draw upon this body of work to tackle multiple interrelated aims :
To be explicit about what we do in design study, why we do it, and what it allows us to know.
To consider the ways in which design and research, theory and practice interrelate.
To understand the relationships between the particular and the general, and ensure that we value both kinds of knowledge if we consider them to be legitimate.
To help us make informed decisions about legitimacy itself.
To explore philosophical positions and provide methodological foundations that can change the ways that we approach, create, assess, and use situated knowledge from design study.

\new{
We reposition
design study as a rich, subjective, and interpretative approach to visualization research inquiry that can be rigorously applied to wicked real-world problems, to construct valuable new knowledge. 
We argue that when established with rigor, such knowledge complements research undertaken from a positivist perspective, contributing beneficially to the wider discipline.
}
%
%
Core to this repositioning are six preliminary criteria that contribute to rigor in visualization design study research, which should be conducted and reported in ways that are
\vspace{0.05cm}
\\
{\fontfamily{qag}\selectfont{\small{INFORMED}}},
{\fontfamily{qag}\selectfont{\small{REFLEXIVE}}},
{\fontfamily{qag}\selectfont{\small{ABUNDANT}}},
{\fontfamily{qag}\selectfont{\small{PLAUSIBLE}}},
{\fontfamily{qag}\selectfont{\small{RESONANT}}}
{\fontfamily{qag}\selectfont{\small{\&}}}
{\fontfamily{qag}\selectfont{\small{TRANSPARENT}}}.

\vspace{0.1cm}

The criteria are inherently dynamic, contradictory, and imperfect knowledge constructs.
We purposefully scope them narrowly to achieve rigor in design study, and note that they do not completely cover the broader considerations necessary for high-quality research. We speculate, however, that the criteria, and our perspective, may transfer to other methods, approaches, and contexts beyond visualization design study. We hope that they spark dialogue and debate -- about design study, applied visualization research, and the underpinnings of the discipline more broadly -- and so move our field forward in interesting and productive ways.


%% file: _tableEnd.tex




\newcommand{\tabWhtTitle}   [1]{\textbf{\huge{#1}}}
\newcommand{\tabWhtText}    [1]{\textit{\Large{#1}}}
\newcommand{\tabGryTitle}   [1]{\textbf{\huge{#1}}}
\newcommand{\tabCriterion}  [1]{\textbf{\Huge{#1}}}
\newcommand{\tabText}       [1]{#1}

\begin{table*}[htp]
\caption{Criteria, their rationale, and suggested ways in which they might be achieved in rigorous visualization design study. \newline
Concepts from the literature that informed these criteria are listed with citations.\vspace{0.5cm}}
\label{tab:criteria}
\resizebox{\textwidth}{!}{%

\fontfamily{qag}\selectfont

\setlength\tabcolsep{18pt}
\renewcommand{\arraystretch}{4.5}
\begin{tabular}{
  m{0.35\textwidth}    
  m{0.50\textwidth}    
  m{0.15\textwidth}    
}

\rowcolor[HTML]{EFEFEF}
\vspace{0.25em}
{\tabCriterion{INFORMED}}
\vspace{1.75em}
& \multicolumn{2}{l}{\tabGryTitle{
Existing knowledge informs design and facilitates new interpretations.
}}

\\
\multicolumn{1}{r}{\tabWhtTitle{Important as}}
& \multicolumn{2}{l}{\tabWhtText{Research informed by existing knowledge supports interesting and meaningful insights and interpretations.}} \\

\multicolumn{1}{r}{\tabWhtTitle{Achievable by}} &
\multicolumn{2}{l}{\tabWhtText{
\shortstack[l]{
\\
using text books, academic papers and technical documentation to develop knowledge of visualization and the domain in which you are working;
\\
seeking opportunities to broaden your exposure to a range of design ideas
}
}
} \\

\multicolumn{1}{r}{\tabWhtTitle{Informed by}} &
\multicolumn{2}{l}{\tabWhtText{
  rich-rigor \cite{tracy10};
  theory-ingrained artefacts \cite{sein11}
}
} \\


\rowcolor[HTML]{EFEFEF}
\vspace{0.25em}
\tabCriterion{REFLEXIVE}
\vspace{1.75em}
& \multicolumn{2}{l}{\tabGryTitle{We affect the research, and the research affects us.
}}

\\
\multicolumn{1}{r}{\tabWhtTitle{Important as}}
& \multicolumn{2}{l}{\tabWhtText{
Reflexivity accounts for, and leverages the inherent subjectivity of design study researchers.}} \\
\multicolumn{1}{r}{\tabWhtTitle{Achievable by}} &

\multicolumn{2}{l}{\tabWhtText{
\shortstack[l]{
making reflexive note-taking core to research and design activity;
striving to recognize and record moments of learning;\\
adopting a structured approach to reflective questioning;\\
finding opportunities and allocating time for critique, and using critical feedback to develop your interpretations -- 
consider using autoethnography,\\
critical friends,
design critique,
structured presentations of work and ideas in progress,
engagement with the rebuttal processes involved in peer review
}
}
} \\
\multicolumn{1}{r}{\tabWhtTitle{Informed by}} &
\multicolumn{2}{l}{\tabWhtText{
  confirmability \cite{lincoln85};
  sincerity \cite{tracy10};
  mutually influential roles, guided-emergence \cite{sein11}
}
} \\



\rowcolor[HTML]{EFEFEF}
\vspace{0.25em}
{\tabCriterion{ABUNDANT}}
\vspace{1.75em}
& \multicolumn{2}{l}{\tabGryTitle{More is better.}}
\\
\multicolumn{1}{r}{\tabWhtTitle{Important as}}
& \multicolumn{2}{l}{\tabWhtText{
Abundance provides opportunities to uncover, relate, understand, and justify meaningful insights.}} \\
\multicolumn{1}{r}{\tabWhtTitle{Achievable by}} &
\multicolumn{2}{l}{\tabWhtText{
\shortstack[l]{
soliciting diverse voices and varied perspectives;
seeking and explaining saturation;\\
developing thick descriptions of observations and interpretations;
investing in long-term, sustained collaboration;\\
using participatory and co-design methods, such as creative visualization-opportunity workshops;
\\
prototyping rapidly and in parallel;
explaining and recording as you design with literate visualization
}
}
} \\

\multicolumn{1}{r}{\tabWhtTitle{Informed by}} &
\multicolumn{2}{l}{\tabWhtText{
  credibility \cite{lincoln85};
  rich-rigor, credibility \cite{tracy10};
  authentic \& concurrent evaluation \cite{sein11}
}} \\


\rowcolor[HTML]{EFEFEF}
\vspace{0.25em}
{\tabCriterion{PLAUSIBLE}}
\vspace{1.75em}
& \multicolumn{2}{l}{\tabGryTitle{
Knowledge claims are evidenced, appropriate, and persuasive.
}
}
\\
\multicolumn{1}{r}{\tabWhtTitle{Important as}}
& \multicolumn{2}{l}{\tabWhtText{
Plausible knowledge claims give others the confidence to use them.
}} \\
\multicolumn{1}{r}{\tabWhtTitle{Achievable by}} &
\multicolumn{2}{l}{\tabWhtText{
\shortstack[l]{
\\
writing memos to develop and capture observations, ideas, and connections;
striving for diversity in analysis and observations;
\\
developing rich accounts of reflective analytic process and reasoning by relating memos to observations and each other;
\\
making knowledge claims explicit and persuasive through sound justifications;
\\
considering crystallization to compare, relate, combine, and link a diverse body of evidence
}
}
}
\\

\multicolumn{1}{r}{\tabWhtTitle{Informed by}} &
\multicolumn{2}{l}{\tabWhtText{
  credibility, dependability \cite{lincoln85};
  rich-rigor, credibility \& meaningful coherence \cite{tracy10}
}} \\


\rowcolor[HTML]{EFEFEF}
\vspace{0.25em}
{\tabCriterion{RESONANT}}
\vspace{1.75em}
& \multicolumn{2}{l}{\tabGryTitle{The research inspires understanding and invites action.}}

\\
\multicolumn{1}{r}{\tabWhtTitle{Important as}}
& \multicolumn{2}{l}{\tabWhtText{Resonant research impacts others in the world.}} \\
\multicolumn{1}{r}{\tabWhtTitle{Achievable by}} &
\multicolumn{2}{l}{\tabWhtText{
\shortstack[l]{
selecting problems to which others will relate;
providing rich, evocative, and situated details of context in reports through thick description;\\
designing artifacts that invite engagement;\\
aiming to have affect in your communication -- move, educate, and challenge those who receive your message;\\
explaining and demonstrating with narrated videos, data stories, annotated imagery, or annotated portfolios;\\
speculating on opportunities  for transfer
}}
}\\

\multicolumn{1}{r}{\tabWhtTitle{Informed by}} &
\multicolumn{2}{l}{\tabWhtText{
  transferability \cite{lincoln85};
  worthy topic, resonance \cite{tracy10};
}} \\

\rowcolor[HTML]{EFEFEF}
\vspace{0.25em}
{\tabCriterion{TRANSPARENT}}
\vspace{1.75em}
& \multicolumn{2}{l}{\tabGryTitle{The reporting invites scrutiny.}}
\\
\multicolumn{1}{r}{\tabWhtTitle{Important as}}
& \multicolumn{2}{l}{\tabWhtText{Transparent descriptions of activities, processes, evidence and claims allows judgments about the other criteria to be made.}}\\
\multicolumn{1}{r}{\tabWhtTitle{Achievable by}} &
\multicolumn{2}{l}{\tabWhtText{
\shortstack[l]{
creating data and task abstractions that contextualize designs;\\
producing rich, comprehensive reports of the research process that are sincere, frank, and self-critical;\\
reporting fully and reflexively on dead-ends and failures;
using effective, creative, and extensive supplementary materials;\\
developing a thorough, findable, annotated evidence base to support claims 
providing an audit trail
}
}} \\

\multicolumn{1}{r}{\tabWhtTitle{Informed by}} &
\multicolumn{2}{l}{\tabWhtText{
  dependability, confirmability \cite{lincoln85};
  sincerity \cite{tracy10}
}} \\

\end{tabular}%
}

\vspace{0.5em}

\centering\color[HTML]{EFEFEF}\rule{0.8\textwidth}{1pt}

\end{table*}